\documentclass[10pt]{article}

\usepackage[dvips]{graphicx}
\usepackage[english]{babel}
\usepackage[latin1]{inputenc}
\usepackage{amsmath}
\usepackage{eucal}
\usepackage{amssymb}
\usepackage{url}
\usepackage{epsfig}
\usepackage{multirow}
\usepackage{balance}

\usepackage{cite} 
\usepackage{stfloats}  
                  
\begin{document}

\title{Topological Feature Based Classification}
\author{
Leto Peel\\
Advanced Technology Centre\\
BAE Systems\\
Bristol, UK\\
Email: leto.peel@baesystems.com
}

\maketitle

\selectlanguage{english}

\begin{abstract}
There has been a lot of interest in developing algorithms to extract clusters or communities from networks.
This work proposes a method, based on blockmodelling, for leveraging communities and other topological features for use in a predictive classification task. Motivated by the issues faced by the field of community detection and inspired by recent advances in Bayesian topic modelling, the presented model automatically discovers topological features relevant to a given classification task.  In this way, rather than attempting to identify some universal best set of clusters for an undefined goal, the aim is to find the best set of clusters \textit{for a particular purpose}.
 Using this method, topological features can be validated and assessed within a given context by their predictive performance.     

The proposed model differs from other relational and semi-supervised learning models as it identifies topological features to \textit{explain} the classification decision.  In a demonstration on a number of real networks the predictive capability of the topological features are shown to rival the performance of content based relational learners.  Additionally, the model is shown to outperform graph-based semi-supervised methods on directed and approximately bipartite networks.
\end{abstract}

\hspace{5mm}


\linespread{0.9}
\section{Introduction}
\label{sec:Introduction}
Networks are found everywhere, spanning a range of domains including physics, sociology, biology, and computer science.   
Where traditional knowledge discovery methods have focused on data attribute values for learning tasks, the interrelations between data instances have now become an important area of research. This work considers the automatic discovery of topological features \textit{which are relevant to a predictive classification task}.  The work is motivated by some of the issues facing the field of community detection \cite{Givan:community} and inspired by recent advances in Bayesian topic modelling \cite{Blei:LDA}.

An example application domain which has benefited from the use of network structures is fraud detection \cite{cortes:guiltassociation, Hill06buildingan, Chau06detectingfraudulent}.  Typical structural indicators of fraud include: communities of interest \cite{cortes:guiltassociation} (fraudulent entities are closer to other fraudsters), structural equivalence \cite{Hill06buildingan} (aliases of a fraudulent entity link to the same set of entities), and bipartite networks \cite{Chau06detectingfraudulent} (fraudsters generating normal activity using ``accomplice'' entities).  It is the aim of this work to identify these types of topological features to predict the unknown class labels in a partially labelled network.

The topological features which have received the most attention in recent years are communities or modules; groups of highly connected nodes within a globally sparse network. 
The task of community detection  is to identify these groups. 
The motivations for conducting community detection are based on the principles of assortative mixing; highly connected nodes share common properties or attributes. 

\begin{figure}
	\centering
		\epsfig{file=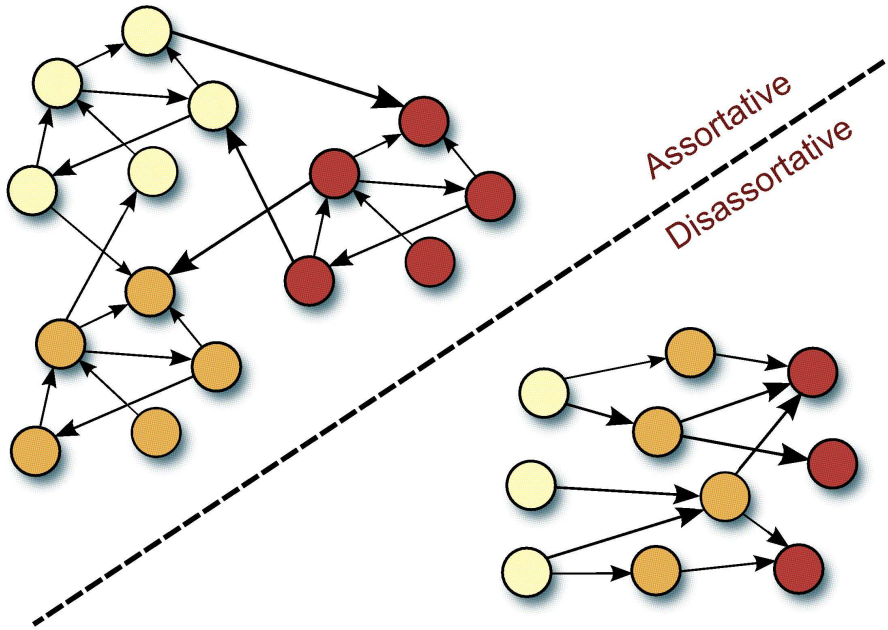, width=\linewidth} 
	\caption{Examples of the topological features (network positions) which are used for classification.  The colours indicate nodes of the same network position. Network positions are defined by their connections to each of the network positions.  Positions may be assortative (top left), tending to link to others of the same position, or disassortative (bottom right), linking to positions different to themselves. network positions. }
	\label{fig:positions}
\end{figure}

Although the problem of community detection may be well known and well studied \cite{Givan:community,Newman:modularity,Palla:overlap,gregory:copra}, it still lacks a formal definition and, in close relation to this, an accepted method for performance assessment.  Common approaches involve either scoring algorithms by the  communities they detect using the Newman-Givan modularity function \cite{Newman:modularity}, or to measure their ability to recover known partitions in the network.  

Modularity measures the difference between the observed linkages within clusters and the expected connectivity of a random graph with the same degree distribution.  However, there are a number of well documented problems with this function such as the resolution limit \cite{fortunato:resolution} and the more recently discovered degeneracy problem \cite{good:performance}.  The degeneracy issue is an important one which provides motivation for this work.  The modularity function has been found to have an exponentially large number of high modularity solutions which are distinctly different from each other.  The implications are that modularity scores alone cannot indicate how good the identified clusters are.  The problem is further compounded by hierarchical and overlapping communities.  Furthermore, modularity scores are not comparable across networks.  

The alternative for performance assessment is to use community detection to recover some known partition by either seeding a \textit{true} partition in a generated network \cite{fortunato-2008} or working with real networks for which a partition is well known; the networks of Sampson's monastery \cite{sampsonmonks} and Zachary's karate club \cite{Zachary} are two popular examples. 
However, in many real networks there are many dimensions in which a ``true'' partition could lie e.g. in a social network, partitions could relate to family relations, business departments, social interest groups, ethnicity, or residential location.  This implies that any known partition of a given network may not be the only valid one and therefore the ability to recover a particular partition which happens to be known seems to be an unfair test for an \textit{unsupervised} method.  The multi-dimensionality of human relations (and other complex systems represented by networks) has provided the motivation for algorithms to detect overlapping communities \cite{Palla:overlap,gregory:copra,Airoldi:MMSB}, but validation of these models are difficult without complete knowledge of the data.  

The work presented here, instead of attempting to optimise a function known to be problematic or to solve an essentially supervised problem without any training, the aim is to try to use these structures as features of the network that can be used in a predictive capacity.  In this way, the clusters found are suitable \textit{for a particular purpose}, i.e. the prediction of node class labels.  This addresses the validation problem, as the clusters found can be assessed according to their predictive performance in a given classification task.  

The proposed model is based on blockmodels which have been previously used for community detection, such as in \cite{reichardt_role} where a blockmodelling approach is formulated in such a way that modularity optimisation is special case. Blockmodels
 also allow the discovery of other types of (disassortative) features which characterise the inter-cluster/community relations and therefore allows a wider search space of topological features.  Furthermore blockmodels provide a descriptive summary of the network; the interpretation of which is covered by extensive literature \cite{wassermanFaust, Snyder1979, White1976}.  

The proposed model is demonstrated on citation, blog and word networks to identify features in the network topology which are then used to predict the classification of unlabeled nodes in the network.  These features are referred to as ``network positions'' and a community is special type of position (Figure \ref{fig:positions}).  The performance of the model on citation networks is found to rival other relational learners, without the use of node attributes (word frequencies).  Finally the model is shown to perform well on (approximately) bipartite networks.  

\section{Related Work}
The work presented draws upon previous work from the fields of community detection, blockmodelling and topic modelling.

In the past decade community detection has received a lot of attention from researchers spanning a wide range of domains and has consequently produced a wide range of approaches to the problem; for a comprehensive review of these approaches see \cite{Fortunato:community}.  Although there is no commonly accepted formal definition for community detection, a lot of approaches focus on the optimisation of the Newman-Givan modularity function \cite{Newman:modularity}.  
The Newman-Givan modularity function, $Q$, of a network partition made up of $K$ disjoint communities is given by:
\begin{displaymath}
  Q = \sum_k^K{\left[ \frac{e_k}{I} - \left(\frac{d_k}{2I}\right)^2 \right]},
\end{displaymath}
where $e_k$ is the number of edges in community $k$, $I$ is the number of edges in the network and $d_k$ is the total degree of the nodes in community $k$.  
Commonly used as it is, the modularity function has been found to suffer from a number of issues such as a resolution limit \cite{fortunato:resolution} and degeneracy of solutions \cite{good:performance}.  Rather than seek to find a globally optimum set of clusters, the proposed approach addresses the degeneracy issue by attempting to find clusters which are able to predict the class of a node.  

In part, due to the lack of formal problem definition, community detection lacks a commonly accepted evaluation procedure.  To address this \cite{fortunato-2008} proposes a benchmark graph generator and in later work \cite{Lancichinetti:comparative} goes on to give a comparative analysis of a selection of community detection algorithms.
Recognising that different community detection algorithms produce desirable outputs under different circumstances, the work in \cite{Peel:Fusion10} considered the problem of selecting an appropriate community detection algorithm based on the properties of the network and community structure.  These methods however rely on there being a single \textit{true} partition, which in many cases is unreasonable.  The work presented here instead considers that there may be alternative partitions and instead searches for the partition relevant to a particular context.  

Blockmodels have been used for social and psychometric analysis for decades \cite{LfWh71, Holland1976, wassermanFaust}.  The name refers to the ``blocks'' of zero and non-zero elements that occur in the adjacency matrix when the rows and columns are reordered such that nodes with similar interaction patterns are adjacent. The clusters of nodes which make up these block patterns are known as \textbf{network positions}.  The pattern of interactions between positions provides a summary of the interactions of the network.  
Bayesian formulations of blockmodelling, known as stochastic blockmodelling \cite{nowicki2001eap}, were developed to create blockmodels by automatically assigning nodes to positions.   
Other extensions of the stochastic blockmodel include the Mixed Membership Stochastic Blockmodel (MMSB) \cite{Airoldi:MMSB} which allows nodes to belong to multiple positions and the non-parametric approaches of the Infinite Relational Model \cite{kemp:IRM} and the Latent Feature Relational Model \cite{NIPS2009_0960} which automatically determines the number of positions. 
While these models display statistically elegant solutions, they do so at a computational cost.  Some more recent attempts at scalable probabilistic models of social networks include the Interaction Component Model for communities (ICMc) \cite{Sinkkonen:Component} model, the Simple Social Network using Latent Dirichlet Allocation (SSN-LDA) \cite{Zhang07_ISI} and Marginal Product Mixture Model (MPMM) \cite{DuBois:2010}. These models assign latent classes to each transaction rather than modelling all possible interactions and so provides the greatest benefit when the networks are sparse.  The associated limitations of these approaches are that, either the receivers and senders are generated from disjoint sets of positions, and/or the memberships of nodes to positions are no longer explicitly inferred.  The first case refers to the SSN-LDA and MPMM models where receivers are generated from one set of network positions and the senders from another set.  This makes models inappropriate for analysis of community structures i.e. because a community is a position in which nodes interact with others of the same position.  This requires sender and receiver positions to be the same.  In the second case, applicable to ICMc and MPMM, the effect is that rather than model the probability of an interaction given the positions of the sender and receiver (i.e. $P(i=1|k_1,k_2)$ where $k_1$ and $k_2$ are positions of sender and receiver), these model the probability of a sender-receiver pair of positions given an interaction (i.e. $P(k_1,k_2|i=1)$).  
An added advantage of these approaches
 is that they are capable of modelling networks with weighted links, where previous blockmodel approaches merely modelled the presence or absence of a relation.  

The main inspiration for this work comes from supervised Latent Dirichlet Allocation (sLDA) \cite{blei:sLDA} which, as the name suggests, is a supervised extension of the topic modelling approach LDA \cite{Blei:LDA}. Latent Dirichlet Allocation is a method for clustering a corpus of documents into mixed membership topics.  The sLDA model extends the LDA approach to identify topics in documents which not only best describes the document structures but also to predict a known response variable (i.e. a classification or regression target) associated with each document.  

The proposed model shares some similarity to graph based semi-supervised learning methods such as \cite{Macskassy03asimple, NIPS2003_AA41, He:Graphsemi, Zhu03semi-supervisedlearning} which aim to label the unlabeled nodes of a partially labelled graph.  However, these methods assume assortativity, i.e. nodes of the same class tend to link to each other, whereas the method herein does not; this is demonstrated in section \ref{sec:classification}

Also related are the collective classification models such as the work of \cite{sen_getoor} and \cite{Kou07stackedgraphical}.  Collective classification is a type of relational learning which uses object attributes together with information about the links between objects to predict class labels.  The work presented here does not use content information and relies only on the link structure of the network.  

\section{Proposed Model}

\begin{figure}
	\centering
		\epsfig{file=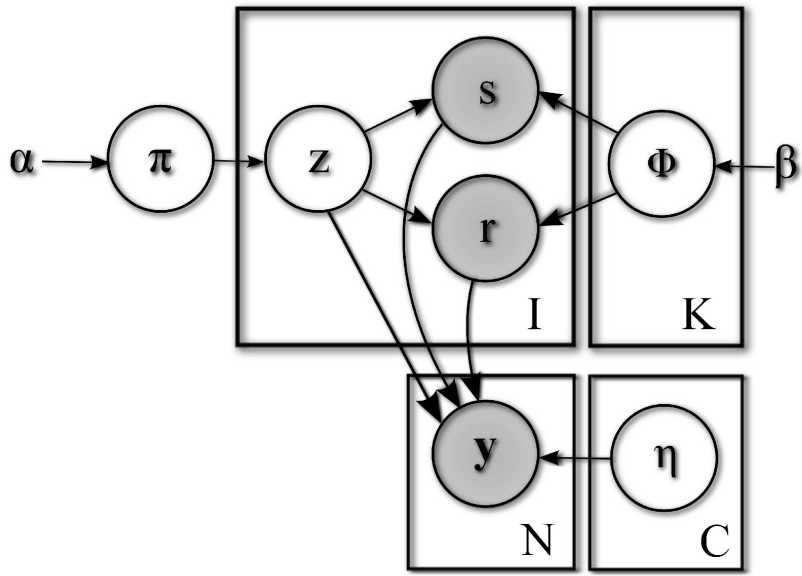, width=.9\linewidth} 
	\caption{A graphical model representation of the supervised blockmodel for sparse networks.  Note that $z$ is a pair of $1$ of $K$ indicator variables $(z_s,z_r)$ drawn from a multinomial of length $K^2$}
	\label{fig:ssMMSBmodel}
\end{figure}

Traditional blockmodelling is based on the idea that a set of latent features, referred to as network positions, are responsible for the interaction patterns between nodes in a network.  These positions represent groups of nodes which have similar interaction patterns.  Nodes of one position link to other nodes of another position with the same probability.  If nodes of a particular position prefer to link to other nodes of the same position, then this position represents a community.  The early approaches to blockmodelling involved manually populating the positions using attribute values to group entities.  More recent contributions \cite{nowicki2001eap,kemp:IRM,Airoldi:MMSB} use probabilistic inference to determine the positions and validate the models against node attributes.   

The proposed model lies somewhere in the middle of these approaches as it infers position membership based on the network structure, but uses class label information to coerce the position assignment such that the positions are relevant to the classification problem.  The model allows multiple network positions to make up a class (node label), or for a particular network position to be incorporated in more than one class.  The model also allows for nodes to take on different positions for different interactions.  
The model is formulated by combining the sLDA model of \cite{wangbleifeifei08} to an adaptation of the ICMc model given in \cite{Parkkinen_Sinkkonen_Gyenge_Kaski_2009} to provide a supervised model which is scalable to large, but sparse networks.

The proposed Supervised Blockmodel for Sparse Networks (SBSN) assumes the following generative process:
\begin{enumerate}
	\item For a given network draw a distribution over the possible $K^2$ network position interactions \footnote{Conceptually, it is best to think of $\mathbf{\pi}$ as a matrix with dimensions $K \times K$ such that each element $\pi_{k_1,k_2}$ represents the probability of observing a link from $k_1$ to $k_2$.} $\pi \sim Dirichlet(\alpha)$
	\item For each position $k \in \{1,2,...,K\}$:
	\begin{enumerate}
		\item Draw a distribution over nodes $\phi \sim$ $ Dirichlet(\beta) $
	\end{enumerate}
	\item For each interaction $i \in \{1,2,...,I\}$:
	\begin{enumerate}
		\item Draw a position interaction pair $z_i = (z_s,z_r)_i$, $z_i \sim Multinomial(\pi)$
		\item Draw a sender node $s_i \sim Multinomial(\phi_{z_s})$
		\item Draw a receiver node $r_i \sim Multinomial(\phi_{z_r})$
	\end{enumerate}
	\item For each node $v \in \{1,2,...,N\} $:
	\begin{enumerate}
		\item Draw a class label $y_v \sim Softmax(\eta,\bar{z}_v)$ where $\bar{z}_v = \frac{1}{n_v}\sum_i{z_{s_i}\delta_{s_i,v} + z_{r_i}\delta_{r_i,v}}$ and $z_{s_i}$ and $z_{r_i}$ are the indicator vectors of length $K$ describing the network position of the sender and receiver nodes  in interaction $i$,  $\delta_{\cdot,\cdot}$ is the Kronecker delta.   $\bar{z}_v$ therefore represents the empirical behavior class frequencies for node $v$.  The softmax function provides the following distribution:
		
			$\quad p(y_v|\eta,\bar{z}_v) = \exp(\eta_{y_v}^T\bar{z}_v)/\sum_c{\exp(\eta_c^T\bar{z}_v)}$
	\end{enumerate}
\end{enumerate}
\hspace{5mm}

Inference of the network positions, $\mathbf{z_s}$ and $\mathbf{z_r}$, and estimation of the model parameters,  $(\mathbf{\pi}, \mathbf{\phi})$ \footnote{The hyperparameters $\alpha$ and $\beta$ are fixed.}  are undertaken using an approximation of the Expectation-Maximisation (EM) algorithm, variational Bayesian EM (VBEM) \cite{Bernardo03thevariational}.
To start with the intractable likelihood is lower bounded using an arbitrary distribution $q$ and Jensen's inequality to give the variational free energy $\mathcal{F}(q,\Theta)$ :
\begin{align}
	\log p&(\mathbf{s,r,y}|\Theta) \notag\\
	&= \log \int p(\mathbf{z,\phi,\pi,s,r,y}|\Theta) d\mathbf{\pi} d\mathbf{\phi} d\mathbf{z} \notag\\
	&= \log \int q(\mathbf{z,\phi,\pi}) \frac{p(\mathbf{z,\phi,\pi,s,r,y}|\Theta) } 
	{q(\mathbf{z,\phi,\pi})} d\mathbf{\pi} d\mathbf{\phi} d\mathbf{z} \notag\\
	&\geq  \int q(\mathbf{z,\phi,\pi}) \log \frac{ p(\mathbf{z,\phi,\pi,s,r,y}|\Theta) } 
	{q(\mathbf{z,\phi,\pi})} d\mathbf{\pi} d\mathbf{\phi} d\mathbf{z}\notag\\
	& = E_q[\log p(\mathbf{z,\phi,\pi,s,r,y}|\Theta)] - E_q[q(\mathbf{z,\phi,\pi})]\notag\\
	& = \mathcal{F}(q,\Theta). \label{freeE}
\end{align}
In regular EM $\mathcal{F}(q,\Theta)$ is optimised with respect to the $q$ distribution (E-step) and the model parameters (M-step) iteratively.  In the E-step the lower bound is saturated by setting $q$ equal to the posterior distribution over latent variables (positions).  However in this case, as computing the posterior is intractable, the posterior is approximated the with the fully factored distribution:
\begin{displaymath}
  q(\mathbf{z,\phi,\pi}) = Dir(\pi_{k_1,k_2}|\omega)\prod_i^I{q(z_i|\lambda_i)}\prod_{k}^K{Dir(\phi_k|\zeta)},
\end{displaymath}
where $\lambda$, $\zeta$ and $\omega$ are the variational parameters of the approximated posterior.  For brevity the full derivation of the variational E and M steps is omitted and only the update functions are included here.  The update steps were derived by closely following the procedure outlined in \cite{wangbleifeifei08} and uses the same approximations and identities.  The interested reader may wish to refer to the appendix in \cite{Airoldi:MMSB} for a full derivation of the variational EM algorithm for the MMSB or \cite{blei:sLDA} for further details of Supervised LDA.  

\subsection{Variational E-Step}
\label{sec:estep}
Inference of the network positions is achieved by optimising the variational free energy $\mathcal{F}(q,\Theta)$ in \eqref{freeE} with respect to the approximating $q$ distribution.
\begin{displaymath}
\begin{aligned}
	\lambda_{i,k_1,k_2} \propto &  \exp \left(  \Psi \left( \zeta_{k_1,s_i} \right) +  \Psi \left( \zeta_{k_2,r_i} \right) + \Psi \left(\omega_{k_1,k_2}\right)  \right.\\
  & \left. + \frac{\eta_{y_{s_i},k_1}}{n_{s_i}}-\frac{h_{i,s_i,k_1}}{h_{i,s_i}^T\lambda_{s_i}^{old}} +   \frac{\eta_{y_{r_i},k_2}}{n_{r_i}}-\frac{h_{i,r_i,k_2}}{h_{i,r_i}^T\lambda_{r_i}^{old}} \right),
\end{aligned}
\end{displaymath}
where $\Psi(\cdot)$ is the derivative of the log Gamma function, $n_{v}$ is the number of times node $v$ is involved in an interaction and $\lambda_{v}$ is a $K$-length vector representing the marginal probability of sender or receiver positions, i.e.:
\begin{displaymath}
\begin{aligned}
 \lambda_{s_i} = \left[\sum_{k_2}{\lambda_{i,1,k_2}}, \sum_{k_2}{\lambda_{i,2,k_2}}, \cdots, \sum_{k_2}{\lambda_{i,K,k_2}}\right]^T, \\
 \lambda_{r_i} = \left[\sum_{k_1}{\lambda_{i,k_1,1}}, \sum_{k_1}{\lambda_{i,k_1,2}}, \cdots, \sum_{k_1}{\lambda_{i,k_1,K}}\right]^T, 
\end{aligned}
\end{displaymath}
and $h_{i,v}^T\lambda_{v}$ represents an approximation to the expectation under the $q$ distribution of the normalising function of the softmax distribution for a node $v$.  This $h_{i,v}$ is given by:
\begin{displaymath}
\begin{aligned}
	h_{i,v} = \sum_c{\exp\left(\frac{\eta_c}{n_v}\right)\prod_{\substack{j\neq i:\\ v\in \left\{s_j,r_j\right\}}}{\left(\sum_k{\lambda_{v,k}\exp\left(\frac{\eta_{c,k}}{n_v}\right)}\right)}}.
\end{aligned}
\end{displaymath}

\subsection{Variational M-Step}
The variational M-step updates the posterior distribution over the model parameters which maximise the variational free energy $\mathcal{F}(q,\Theta)$ for the current distribution over the latent variables $\mathbf{z}$. 
\begin{displaymath}
\begin{aligned}
  &\zeta_{k_1,v} = \sum_{i=1}^I{\delta{s_i,v}\sum_{k_2}{\lambda_{i,k_1,k_2}} + \delta{r_i,v}\sum_{k_2}{\lambda_{i,k_2,k_1}} + \beta}, \\
  &\omega_{k_1,k_2} = \sum_{i=1}^I{\lambda_{i,k_1,k_2} + \alpha},\\
\end{aligned}
\end{displaymath}
$\eta$ is found using conjugate gradient to optimise the free energy terms of \eqref{freeE} corresponding to $\eta$: 
\begin{displaymath}
\begin{aligned}
	\mathcal{F}_{[\eta_{1:C}]} =& \sum_{v}{\eta_{y_{v}}^T\bar{\lambda}_v - \log \sum_c{\prod_{\substack{i:v\in\\ \{s_i,r_i\}}}{\left(\sum_k{\lambda_{v,k}\exp\left(\frac{\eta_{c,k}}{n_v}\right)}\right)}}},\\
\end{aligned}
\end{displaymath}
where $\bar{\lambda}_v = \frac{1}{n_v}\sum_i{\lambda_{s_i}\delta_{s_i,v} + \lambda_{r_i}\delta_{r_i,v}}$.  Conjugate gradient requires the following derivatives:
\begin{displaymath}
\begin{aligned}
  \frac{\partial \mathcal{F}_{[\eta_{1:C}]}}{\partial \eta_{c,k}} =& \sum_v{\bar{\lambda}_v\delta_{y_{v},c}} \\
  & - \sum_v{\frac{\prod_{i:v\in\{s_i,r_i\}}{\left(\sum_l{\lambda_{v,l}\exp\left(\frac{\eta_{c,l}}{n_v}\right)}\right)}}{\sum_c{\prod_{i:v\in\{s_i,r_i\}}{\left(\sum_l{\lambda_{v,l}\exp\left(\frac{\eta_{c,l}}{n_v}\right)}\right)}}}}\\
  & \times \sum_{i:v\in\{s_i,r_i\}}{\frac{\frac{1}{n_v}\lambda_{v,k}\exp\left(\frac{\eta_{c,k}}{n_v}\right)}{\sum_l{\lambda_{v,l}\exp\left(\frac{\eta_{c,l}}{n_v}\right)}}}.
\end{aligned}
\end{displaymath}

\subsection{Prediction}
Predicting unlabeled nodes requires inference of the network positions given the rest of the network.  As the class label is unknown the inference is performed as in Section \ref{sec:estep} but without the terms involving $\eta$.  Classification of a test node is given by:
\begin{displaymath}
	y_v^* = \arg \max_{y \in \{1,...,C\}} E_q \left[ \eta_y^T \bar{z_v} \right] = \arg \max_{y \in \{1,...,C\}} \eta_y^T \bar{\lambda}_v.
\end{displaymath}

\section{Experimental Results}
This section describes the experiments performed using the SBSN model.  In all experiments the parameters $\alpha$ and $\beta$ were fixed at 2.0 and $\frac{1}{C}$ respectively; these values were found in general to give good classification performance.  The $\lambda$ and $\eta$ parameters were initialised randomly. In each classification experiment the SBSN model was fit to the whole network by iterating the variational E and M steps described in the previous section until convergence of the free energy.  Only a randomly selected proportion of the network nodes in each experiment were labelled (training set), the remaining nodes were used as a test set.  

\subsection{Data}
Four publicly available datasets were used to demonstrate classification using the model.  The first two are the citation networks Cora and Citeseer datasets from \cite{sen:aimag08} which consists of 2708 and 3312 nodes representing papers and 5429 and 4732 edges representing citations respectively.  The Cora dataset contains papers from 7 categories and Citeseer has 6.   
The third is the AGBlog dataset, the largest connected component from the graph of the political blog dataset found in \cite{Adamic05thepolitical}.  This network has 1222 nodes labeled ``Liberal'' or ``Conservative'' and 19021 edges connecting them.  The fourth dataset is a word network from \cite{Newman:words} comprised of the 112 most frequently occurring nouns and adjectives in the novel \textit{David Copperfield} by Charles Dickens.  This network contains a link between words whenever they appear adjacent to each other.  It is approximately bipartite and is used to demonstrate the SBSN model's ability to extract disassortative features.

\subsection{Classification}
\label{sec:classification}
Experiments were run to investigate the effect of various factors on the classification performance; the maximum number of positions (K), initialisation of $\phi$, and the proportion of the network that was labelled (training set size).  Performance is measured according to the Macro-averaged F1 measure given by:

\begin{displaymath}
	F_1 = \frac{2TP}{2TP + FN + FP},
\end{displaymath}
where TP, FN, and FP correspond to the true positive, false negative and false positive rates respectively.  The F1 measure represents the harmonic mean of the precision and recall values.  For the multi-class problems the macro-average is used - i.e. the F1 score is calculated for each class and then averaged.  Each experiment was run 25 times and the performance scores reported reflect the average over these runs.

\begin{figure}
	\centering
		\epsfig{file=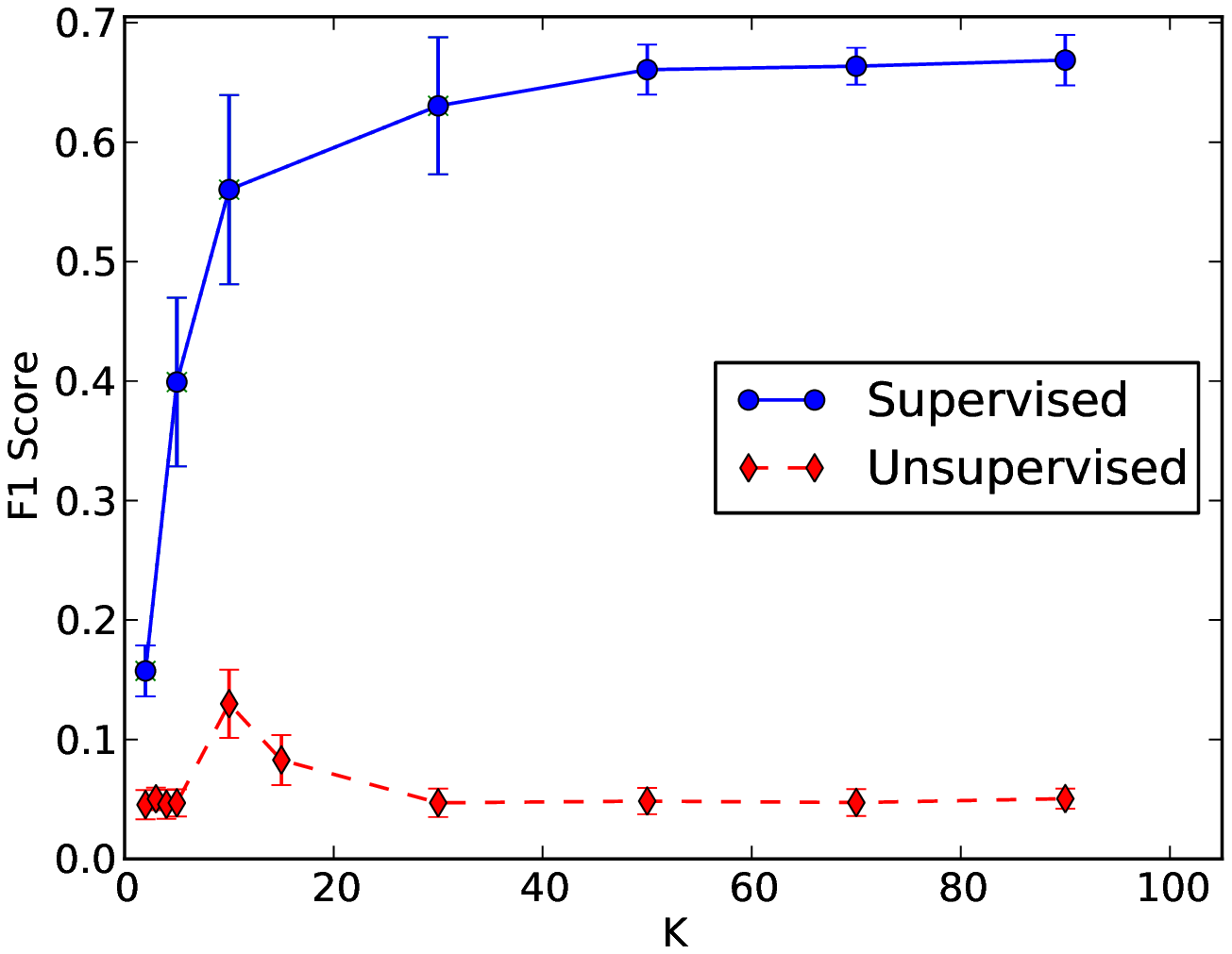, width=.47\linewidth, clip=true, trim=33 15 40 28}
		\epsfig{file=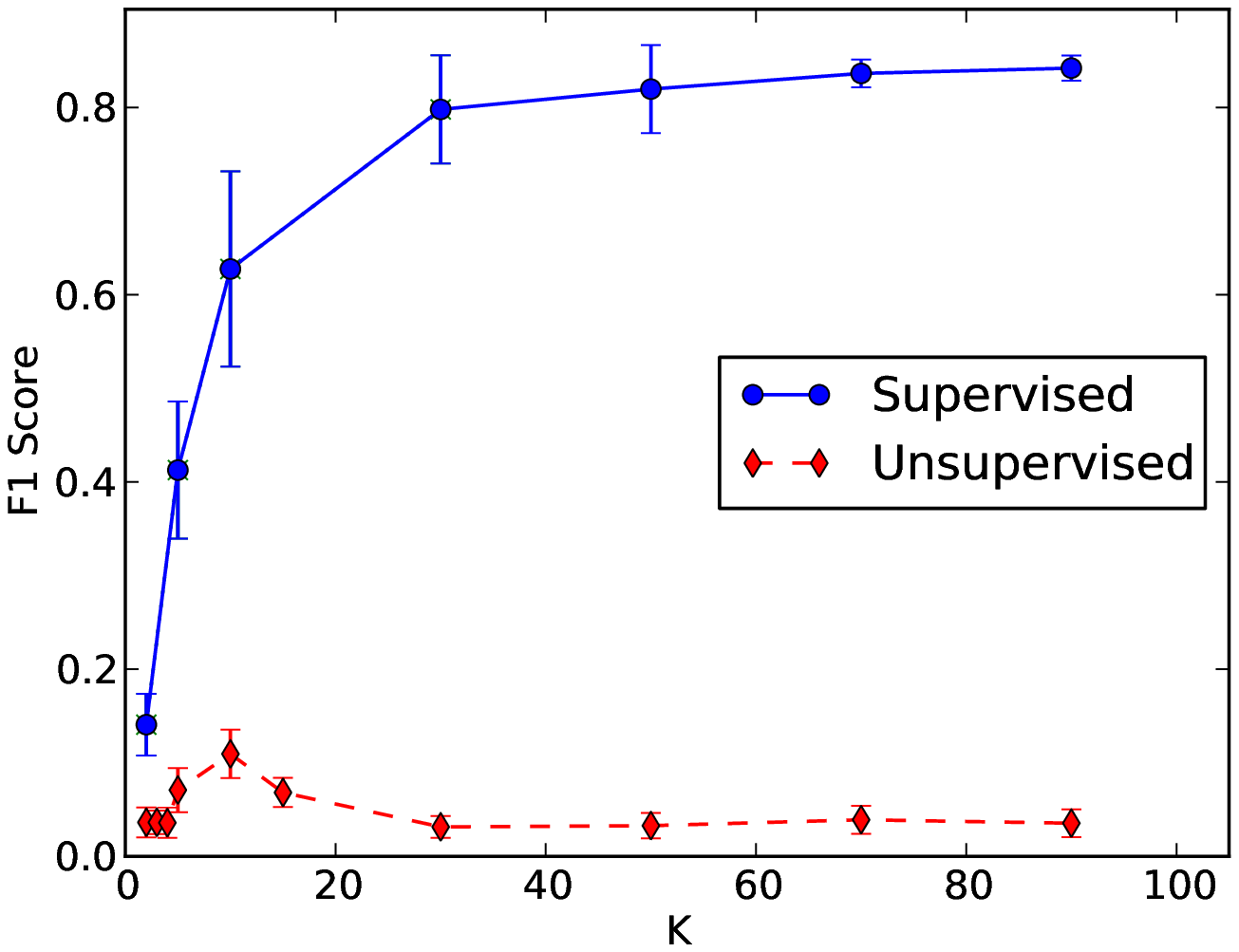, width=.47\linewidth, clip=true, trim=33 15 40 28}
	\caption{Macro-F1 scores for the Citeseer (left) and Cora (right) dataset as the maximum number of positions (K) is varied.  The ``Supervised'' data points show the performance of the SBSN model compared against the baseline ``Unsupervised'' version which is trained without using class labels to infer the positions.}
	\label{fig:corak}
\end{figure}

Figure \ref{fig:corak} shows the classification performance on the Citeseer and Cora datasets for different values of the model parameter K, the maximum number of positions.  As a baseline, the performance of the unsupervised version of the model is also shown (i.e. the terms containing node labels were omitted in the update described in section \ref{sec:estep}).  Each data point represents the mean of 25 tests and the error bars represent one standard deviation.  For each run, two thirds of the network were randomly selected for training and the remaining nodes used for testing.  It can be seen that the performance improves as K increases, however on closer inspection it was found that often the membership for a lot of these positions tended to zero (due to the regularisation effect of the hyperparameters).  It is curious then to see that increasing K improves the performance.  It is suggested that this could be that a greater number of initial positions allows greater freedom for position assignments to change.  

Table \ref{cite_perf} compares the prediction performances on the Citeseer and Cora datasets with the published results of the best collective classifiers in \cite{sen_getoor} and \cite{Kou07stackedgraphical}.  Collective classification is a type of relational learning which uses object attributes together with information about the links between objects to predict class labels.  The results of the SBSN model demonstrate comparable prediction performance rather than consistently outperforming the collective classification models.  The interesting result here is that while the collective classification models use attribute information (i.e. the word frequencies of the documents), the SBSN model does not.  This shows that, for these datasets, the link structure alone is indicative of the class and \textit{suggests} that beyond the topology the content information is redundant. 

\begin{figure}
	\centering
		\epsfig{file=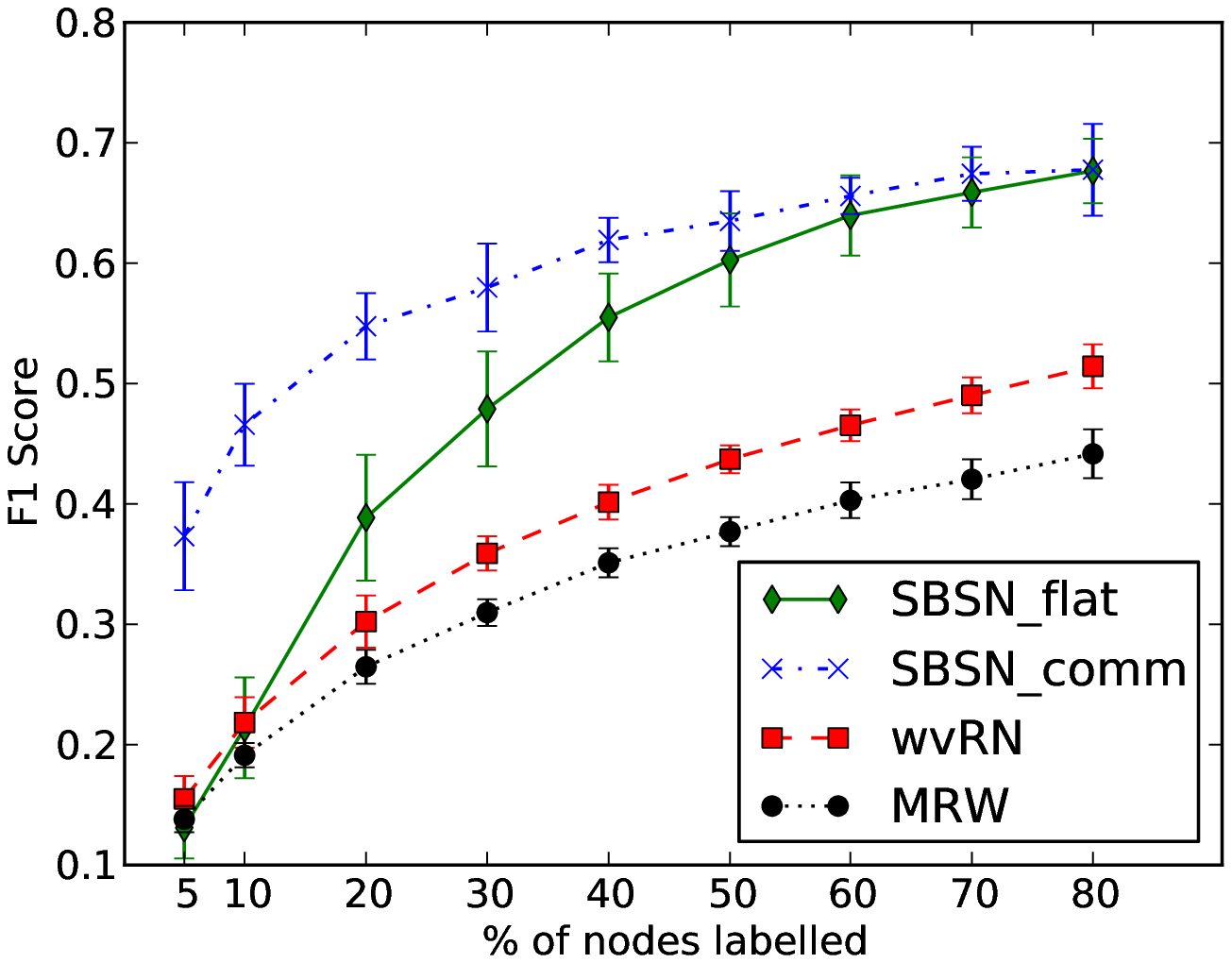, width=.47\linewidth, clip=true, trim=33 15 40 27} 
		\epsfig{file=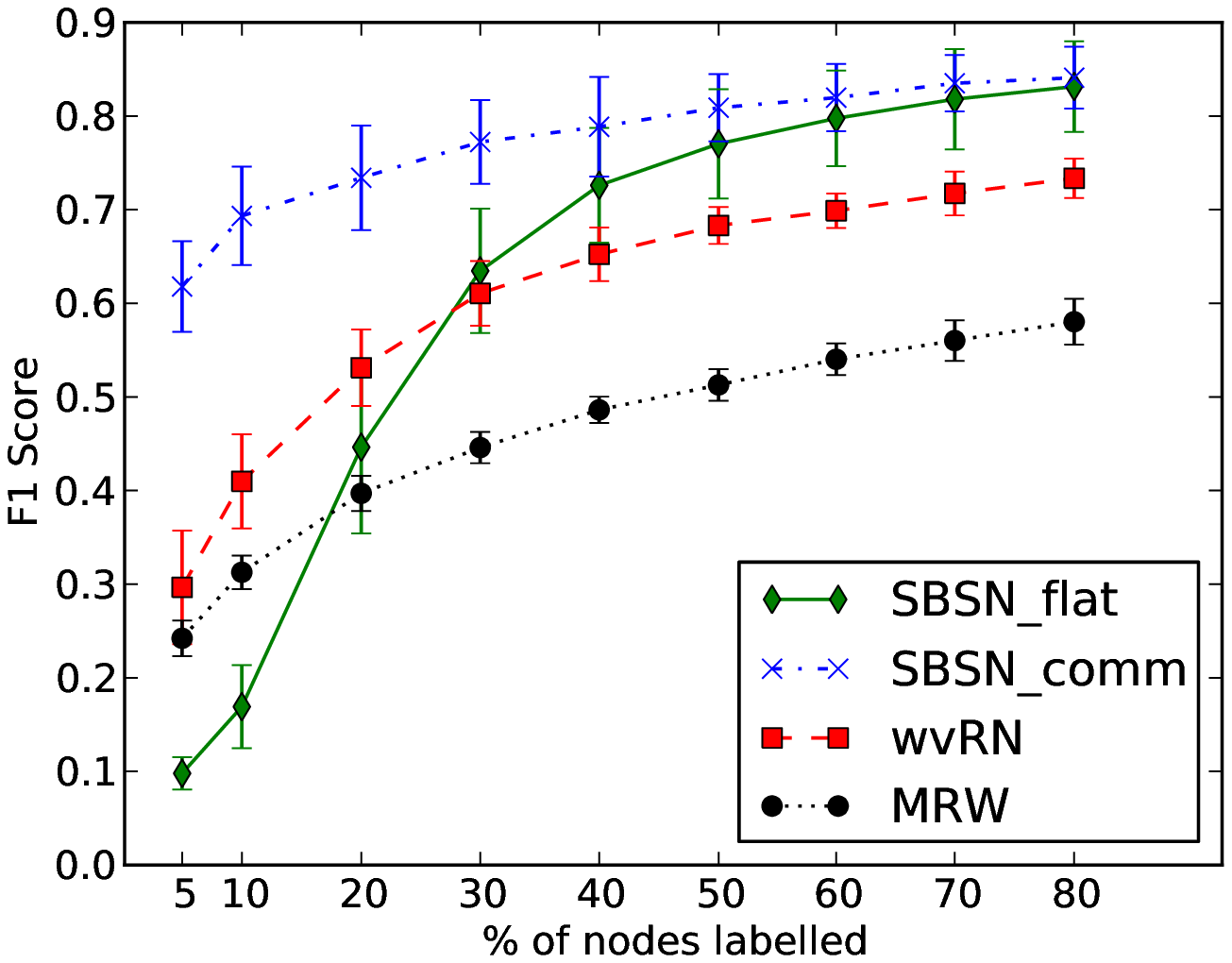, width=.47\linewidth, clip=true, trim=33 15 40 27}
	\caption{Macro-F1 scores for the Citeseer (left) and Cora (right) datasets for different training set sizes.  A comparison is made of the performance of naive initialisation of $\pi$ (flat) and the performance a ``community''-based initialisation (comm).  Also included are two semi-supervised algorithms wvRN and MRW.}
	\label{fig:citesamp}
\end{figure}

\begin{figure}
	\centering
	\epsfig{file=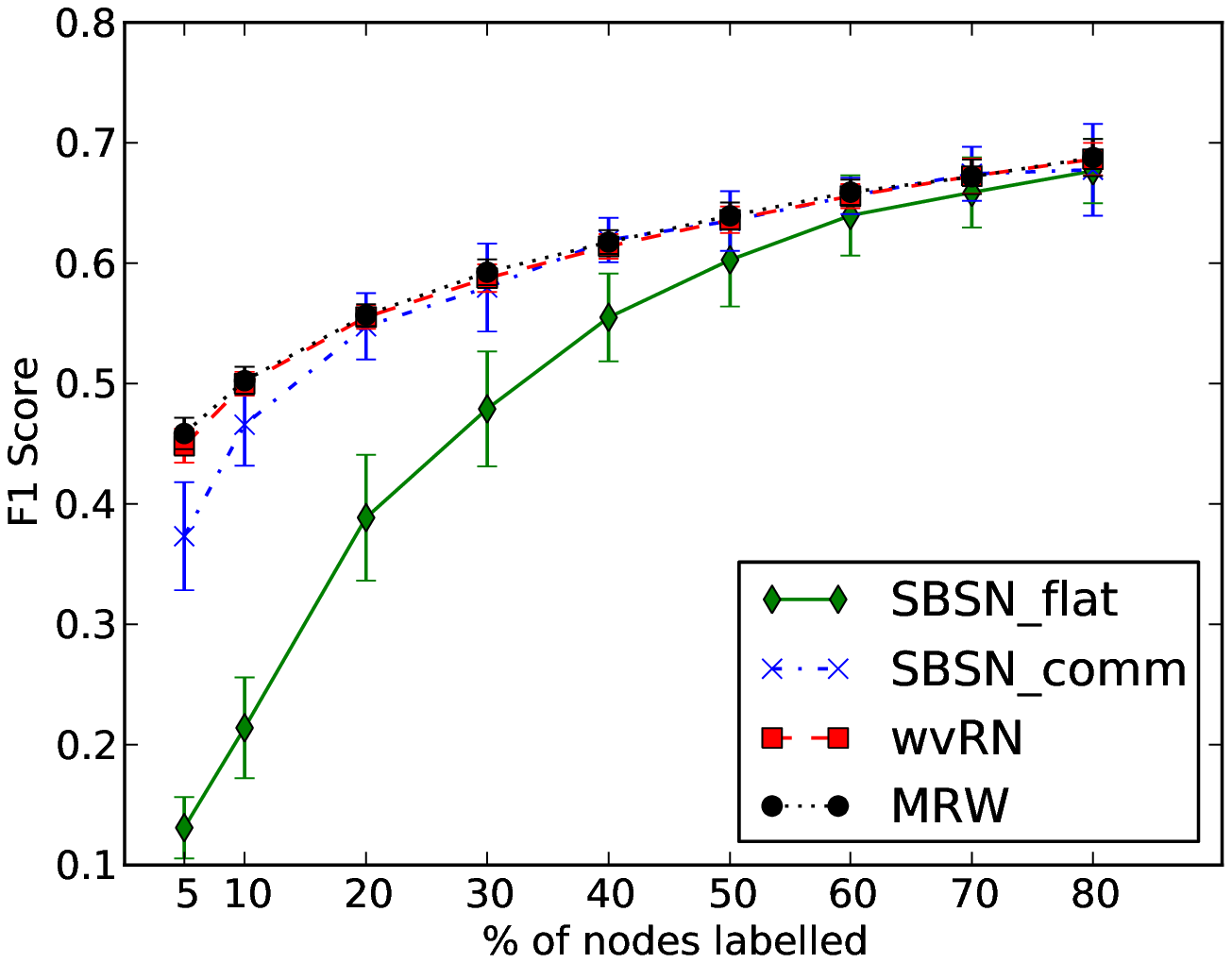, width=.47\linewidth, clip=true, trim=33 15 40 27}
		\epsfig{file=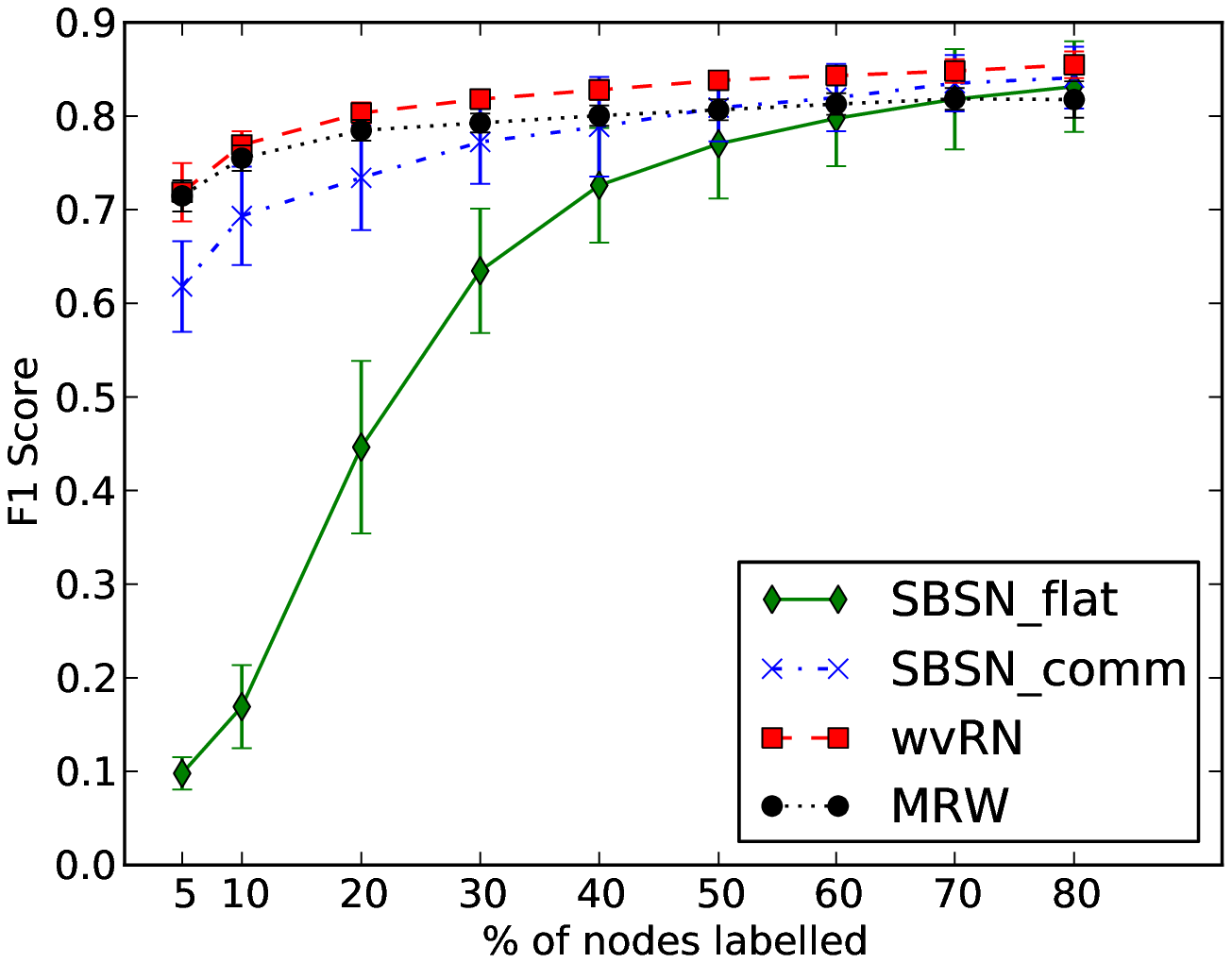, width=.47\linewidth, clip=true, trim=33 15 40 27}
	\caption{Macro-F1 scores for the Citeseer (left) and Cora (right) datasets for different training set sizes.  Similar to figure \ref{fig:citesamp}, but here the algorithms wvRN and MRW are run on undirected versions of the networks.}
	\label{fig:corasamp}
\end{figure}

\begin{figure}
	\centering
		\epsfig{file=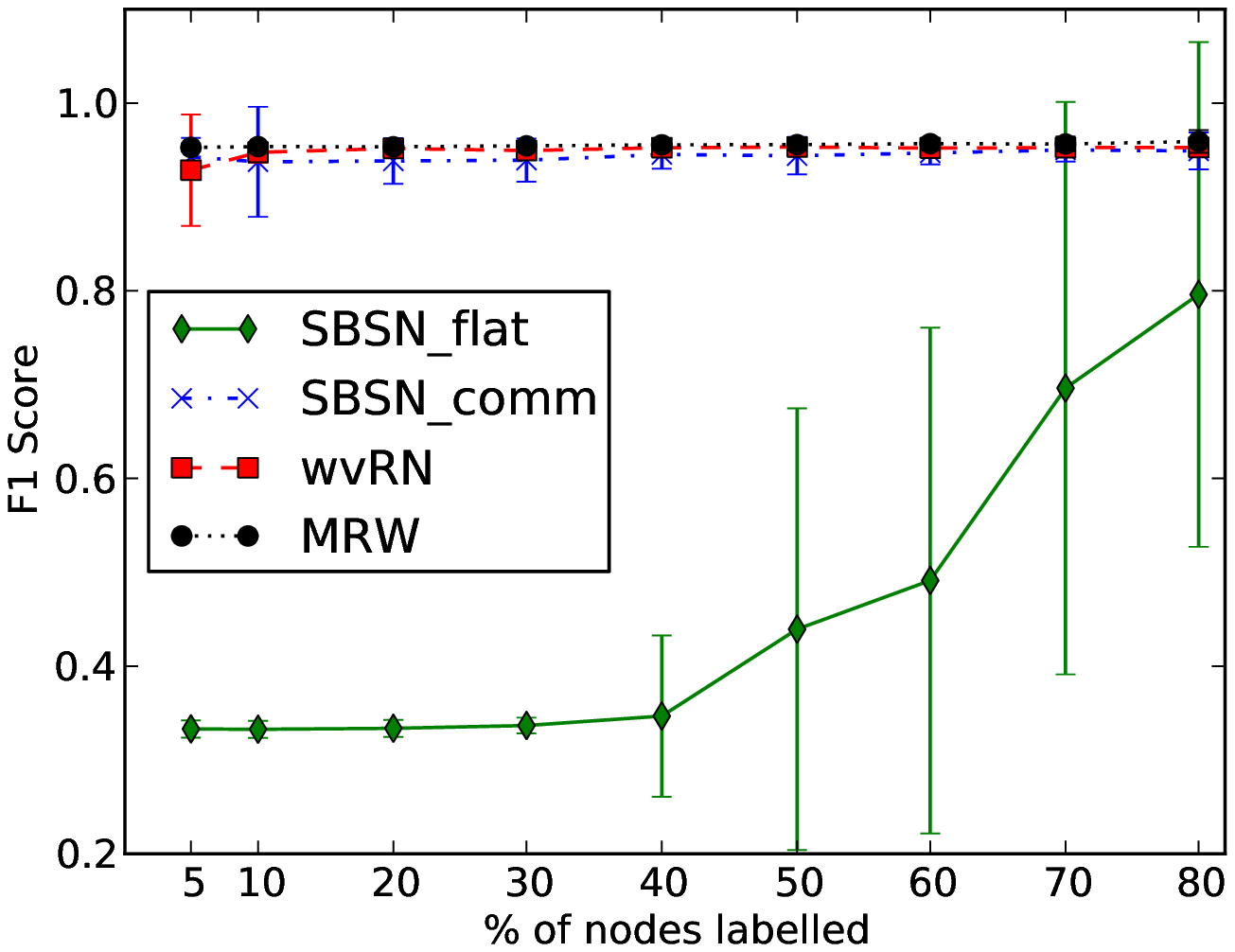, width=.47\linewidth,  clip=true, trim=33 15 40 27}
	\caption{Macro-F1 scores for the Agblog dataset for different training set sizes.  A comparison is made of the performance of naive initialisation of $\pi$ (flat) and the performance of ``community''-based initialisation (comm).}
	\label{fig:agblogsamp}
\end{figure}

Figures \ref{fig:citesamp}--\ref{fig:agblogsamp} show the prediction performance on the citation and blog networks as training set size varies (from 5\% to 80\%).  Two semi-supervised methods, weighted vote Relational Neighbour classifier (wvRN) \cite{Macskassy03asimple} and Multi-Rank Walk (MRW) \cite{lin:semi}, were also run on the same datasets to provide a comparison\footnote{\noindent Using code downloaded from \\ http://www.cs.cmu.edu/$\sim$frank/code/asonam2010-code.zip}.  In comparing the SBSN model with the semi-supervised learners, it was found that the performances were surprisingly low (Figure \ref{fig:citesamp}).  It was found that this was due to the networks being directed and that these algorithms were not designed for directed networks.  The experiments were re-run using undirected versions of these networks which were constructed by repeating each link in the opposite direction (Figure \ref{fig:corasamp}).  It can be seen that the performance of the SBSN model (labelled SBSN\_flat) degrades substantially as the proportion of labelled nodes decreases.  This is because of the large search space of possible position assignments relative to the amount of data.  To compensate for this, instead of running a single M-step to intialise the distribution over $\pi$, the distribution was initialised to favour assortative positions, i.e.:

\begin{displaymath}
	\omega = \begin{bmatrix}
       \frac{I}{K}+\alpha & \alpha & \cdots &\alpha           \\[0.3em]
       \alpha & \frac{I}{K}+\alpha & \cdots & \alpha \\[0.3em]
       \vdots & \vdots & \ddots&  \\[0.3em]
       \alpha & \alpha & & \frac{I}{K}+\alpha
     \end{bmatrix}.
\end{displaymath}

This initialisation, takes into account the assumption made by the semi-supervised algorithms that nodes link to other nodes of the same type (assortativity).  This results in the scores labelled SBSN\_comm.  

\begin{figure}
	\centering
		\epsfig{file=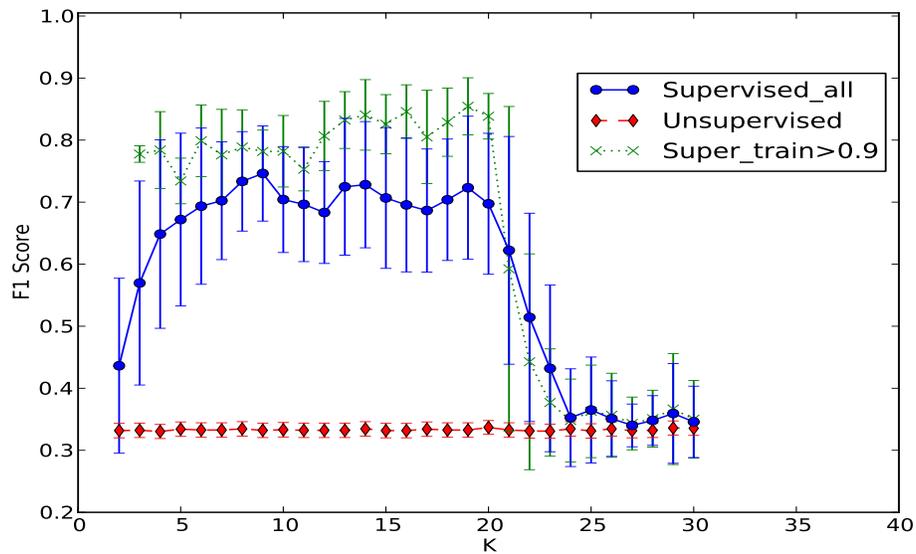, width=\linewidth, height=.6\linewidth, clip=true, trim=35 15 45 38}
	\caption{Macro-F1 scores for the Words dataset as the maximum number of positions (K) is varied. Here a similar comparison is drawn between ``Supervised'' and ``Unsupervised'' versions.  It can be seen that there is high variance in the performance of the SBSN model due to occasions where the model has failed to fit the data.  For comparison the F1 scores are shown for all the models which achieved an F1 score $>$ 0.9 on the training set. }
	\label{fig:wordk}
\end{figure}

Figure \ref{fig:wordk} shows performance of the SBSN on the classification of adjectives and nouns in the Words dataset. The graph shows that the model performs well with a maximum number of positions ($K$) between 5 and 20.  For larger values of $K$ the performance degrades due to overfitting of the training data.  It can be seen that there is high variance in the performance of the SBSN model. Closer inspection of the results revealed a bimodal distribution for the prediction performance which was due to occasions where the model failed to fit the training data, i.e. the classification performance on the training set was poor (a similar situation was observed on the Agblog results shown in Figure \ref{fig:agblogsamp}).  To allow for this Figure \ref{fig:wordk} also shows the performance of models which fit the training set well based on a classification performance on the training set data of at least 0.9.  Table \ref{word_perf} gives the performance scores of the SBSN model in comparison to the semi-supervised models, showing that the SBSN model is able to deal with disassortative classes while these semi-supervised models cannot.

\begin{table}[!t]
\caption{Classification performance --- SBSN vs. Relational Learners}
\label{cite_perf}
\centering
\begin{tabular}{|c|c|c|c|c|} 
\hline & \multicolumn{2}{|c|}{Citeseer network} & \multicolumn{2}{|c|}{Cora network}\\
\hline  & F1  & Accuracy & F1  & Accuracy \\ 
\hline
\hline MF \cite{sen_getoor} & 0.6291 & 0.7267 & 0.7970 & 0.8261\\ 
\hline LBP \cite{sen_getoor} & 0.6264 & 0.7294 & 0.8248 & 0.8449 \\
\hline Stacked \cite{Kou07stackedgraphical} & - & 0.598 & - & 0.739  \\
\hline SBSN & 0.6705 & 0.7029 & 0.8420 & 0.8519 \\
\hline
\end{tabular}
\end{table}

\begin{table}[!t]
\renewcommand{\arraystretch}{1.3}
\caption{Classification performance --- SBSN vs. Semi-supervised}
\label{word_perf}
\centering
\begin{tabular}{|c|c|c|} 
\hline & \multicolumn{2}{|c|}{Words network} \\
\hline  & F1 measure & Accuracy \\ 
\hline
\hline wvRN \cite{Macskassy03asimple} & 0.4216 & 0.5289 \\ 
\hline MRF \cite{lin:semi} & 0.4411 & 0.4614 \\
\hline SBSN & 0.7462 & 0.7484 \\
\hline
\end{tabular}
\end{table}

\subsection{Block Model Analysis}
A property of the SBSN model which sets it apart from the methods of the previous section, is that it identifies the topological features, in the form of network positions, upon which the classification is based.  Using the variable distributions (specifically $\pi$ and $\phi$) of the fitted model, qualitative analysis can be undertaken to understand the network positions and the pattern of links between them which are behind the classification decision.   

\begin{figure}
	\centering
		\epsfig{file=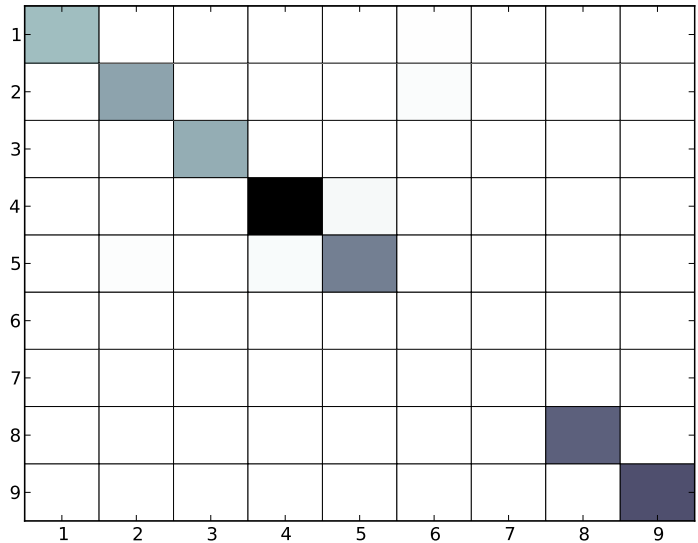, width=.8\linewidth, clip=true, trim=45 30 45 35}
	\caption{Block model image matrix summarising the interaction patterns of the Cora network.  The darker the block the higher the probability of interaction.  It can be seen from the diagonal blocks that the positions found favour interacting with others of the same position}
	\label{fig:allclassblocks}
\end{figure}

To understand the links between positions, a block model matrix is constructed by rearranging the $K^2$ elements of $\pi$ into a matrix such that the rows and columns refer to the positions of the interaction source and target respectively.     Figure \ref{fig:allclassblocks} shows the blockmodel image matrix for the discovered positions in the Cora dataset.  This matrix summarises the interaction patterns in the network and shows the relative probability of an interaction between different network positions (darker shaded blocks indicate higher probability).  The diagonal blocks of the image matrix indicate the presence of assortative groups (communities).  

\begin{figure}
	\centering
		\epsfig{file=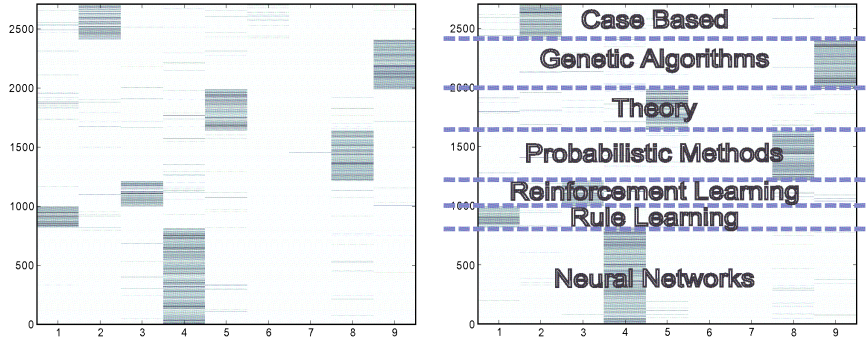, width=\linewidth, height=.43\linewidth}
	\caption{Image matrix of $\phi$ distribution.  Each row describes the membership of the nodes (columns) in the network.  The nodes are ordered by class. The image matrix on the right is the one on the left annotated with class labels.  It can be seen that each class usually corresponds to a particular network position (i.e. Neural Networks are usually position 4)}
	\label{fig:phiblock}
\end{figure}

Visualisation of $\phi$ distribution in a similar way shows which nodes (rows) are members of each of the positions (columns); darker shades indicate a higher level of membership.  From Figure \ref{fig:allclassblocks} it can be seen that positions 6 and 7 appear not to favour interactions with any position.  Cross-referencing this with the image matrix of the $\phi$ distribution in Figure \ref{fig:phiblock} shows that there is only a low level of membership to these positions.  Ordering the rows of the $\phi$ image matrix such that nodes of the same class are adjacent reveals the positions that are indicative of each of the classes; for the Cora dataset this is highlighted in the right hand side of Figure \ref{fig:phiblock} where the image is segmented and annotated with the subject labels.  

\begin{figure*}
	\centering
		\epsfig{file=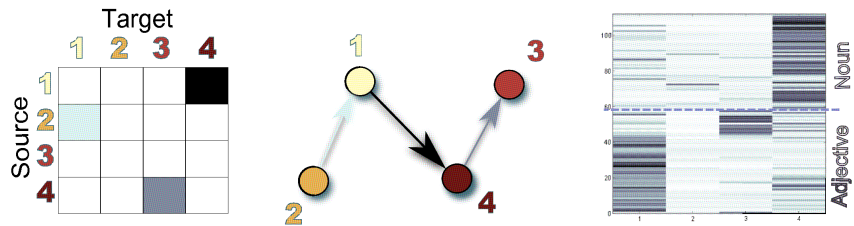, width=.95\linewidth, height=.28\linewidth}
	\caption{Interpretation of the SBSN model applied to the Words dataset.  The blockmodel image matrix (a) describes the probability of any pair of roles interacting, i.e. this is a visual representation of the multinomial parameter $\pi$.  This can be interpreted as a summary of the network interactions (b).  The image matrix of the position memberships ($\phi$) (c) shows which nodes (rows) belong to each of the network positions (columns).  This shows that the Adjectives are usually position 1 and 3 while the Nouns are 2 and 4}
	\label{fig:classblocks}
\end{figure*}

Figures \ref{fig:classblocks}a and \ref{fig:classblocks}c show the $\pi$ and $\phi$ image matrices respectively for the Words dataset.  The off-diagonal blocks of the blockmodel image show the disassortative nature of the positions.  This image can be used to construct a summary network to describe the interactions between positions (Figure \ref{fig:classblocks}b).  In this case the network summary forms a chain.  Cross-referencing with the $\phi$ distribution indicates that the majority of interactions are from the adjectives of position 1 to the nouns of position 4.  This reflects the fact that in the English language adjectives usually precede nouns.

\section{Conclusions and Future Work}
This work has presented the Supervised Blockmodel for Sparse Networks (SBSN), a model for jointly modelling relational and class label information.  The proposed model has been demonstrated to perform well in predictive classification tasks on real world benchmark datasets.  This model differs from other relational and semi-supervised learning models because in addition to classification, it identifies topological features to explain the classification decision; where these features relate to the network positions from blockmodelling.  Initialisation using prior knowledge of the data structure and the rejection of poor fitting models has been considered.  Real world networks in which the SBSN is comparable to the classification performance of relational learners which use additional node attribute information has been demonstrated.  The SBSN has been found to perform similarly to basline semi-supervised learners, although prior assumptions (the same used by the semi-supervised algorithms) are required to obtain good performance when the proportion of labelled data is small.  In addition, the SBSN model can perform well with directed networks and disassortative classes.   

Although not investigated here, the current formulation of SBSN is amenable to weighted networks.  It is left to future work to investigate the multiple separate classification tasks on a single network to explore how the predictive network positions change with the classification task.

\section{Acknowledgments}
This work was undertaken as part of an Engineering
Doctorate at the EngD VEIV Centre for Doctoral Training at
University College London.

\bibliographystyle{ieeetr}
\balance
\bibliography{sigproc} 

\end{document}